# Testing Photometric Techniques for Measuring the Rest-frame UV Spectral Slope against JWST PRISM Spectroscopy

Alexa M. Morales[1,2,21], Steven L. Finkelstein[1,2], Pablo Arrabal Haro[3,22], Micaela B. Bagley[3,4], Antonello Calabrò[5], Óscar A. Chávez Ortiz[2,4], Kelcey Davis[6], Mark Dickinson[7], Eric Gawiser[8], Mauro Giavalisco[9], Nimish P. Hathi[10], Michaela Hirschmann[11], Jeyhan S. Kartaltepe[12], Anton M. Koekemoer[10], Arianna S. Long[13], Ray A. Lucas[10], Fabio Pacucci[14,15], Casey Papovich[16], Borja Pautasso[17], Nor Pirzkal[18], Anthony J. Taylor[1,2], Alexander de la Vega[19], Stephen M. Wilkins[17,20], and L. Y. Aaron Yung[10]

[1] Department of Astronomy, The University of Texas at Austin, 2515 Speedway, Austin, TX 78712, USA; alexa.morales@utexas.edu
[2] Cosmic Frontier Center, The University of Texas at Austin, Austin, TX, USA
[3] Astrophysics Science Division, NASA Goddard Space Flight Center, 8800 Greenbelt Road, Greenbelt, MD 20771, USA
[4] Department of Astronomy, The University of Texas at Austin, Austin, TX, USA
[5] INAF - Osservatorio Astronomico di Roma, via di Frascati 33, 00078 Monte Porzio Catone, Italy
[6] Department of Physics, 196 Auditorium Road, Unit 3046, University of Connecticut, Storrs, CT 06269, USA
[7] NSF's National Optical-Infrared Astronomy Research Laboratory, 950 N. Cherry Avenue, Tucson, AZ 85719, USA
[8] Department of Physics and Astronomy, Rutgers, The State University of New Jersey, Piscataway, NJ 08854, USA
[9] University of Massachusetts Amherst, 710 North Pleasant Street, Amherst, MA 01003-9305, USA
[10] Space Telescope Science Institute, 3700 San Martin Drive, Baltimore, MD 21218, USA
[11] Institute of Physics, Laboratory of Galaxy Evolution, Ecole Polytechnique Fédérale de Lausanne(EPFL), Observatoire de Sauverny, 1290 Versoix, Switzerland
[12] Laboratory for Multiwavelength Astrophysics, School of Physics and Astronomy, Rochester Institute of Technology, 84 Lomb Memorial Drive, Rochester, NY 14623, USA
[13] Department of Astronomy, The University of Washington, Seattle, WA 98195, USA
[14] Center for Astrophysics | Harvard & Smithsonian, 60 Garden Street, Cambridge, MA 02138, USA
[15] Black Hole Initiative, Harvard University, 20 Garden Street, Cambridge, MA 02138, USA
[16] George P. and Cynthia Woods Mitchell Institute for Fundamental Physics and Astronomy, Department of Physics and Astronomy, Texas A&M University, College Station, TX, USA
[17] Astronomy Centre, University of Sussex, Falmer, Brighton BN1 9QH, UK
[18] ESA/AURA Space Telescope Science Institute, 3700 San Martin Drive, Baltimore, MD 21218, USA
[19] Department of Physics & Astronomy, University of California, 900 University Avenue, Riverside, CA 92521, USA
[20] Institute of Space Sciences and Astronomy, University of Malta, Msida MSD 2080, Malta

## Abstract

We present a sample of 53 galaxy spectra at $z_{\rm SPEC} \sim 5$–12 from the JWST CEERS and RUBIES surveys, combining NIRSpec PRISM spectroscopy with NIRCam photometry. We aim to use these data to establish best practices for measuring the UV spectral slope ($\beta$) in the era of JWST. We adopt power-law fits to the rest-frame UV continuum from the spectroscopic data as our fiducial, or "true," $\beta$ values, and compare them to photometric estimates derived through three methods: (1) photometric power-law fitting, (2) power-law fitting to a spectral energy distribution (SED) model fitted to the photometry, (3) single-color fitting near the Lyman break, and (4) single-color fitting at fixed rest-frame wavelengths. We find that photometric power-law fitting and SED fitting are comparable and most closely recover the spectroscopic slopes, with minimal bias and scatter. Single-color estimates, while commonly used in past studies, show the most significant deviations and are not recommended when more than a single color is available. Our results highlight the limitations and strengths of each approach and provide practical guidance for measuring $\beta$ from photometry when spectra are unavailable or are of insufficient quality.

*Unified Astronomy Thesaurus concepts:* Early universe (435); Galaxy formation (595); Galaxy evolution (594); Astronomical methods (1043); Ultraviolet spectroscopy (2284)

## 1. Introduction

When studying the characteristics of active star-forming galaxies and their stellar populations, an analysis of the rest-frame ultraviolet (UV) regime provides insight into the massive stars within these galaxies (D. Calzetti et al. 1994; A. E. Shapley et al. 2003). These stars emit UV radiation, which, in turn, depends on various astrophysical properties. One well-known observable is the rest-frame UV spectral slope, $\beta$. The UV spectral slope measures the steepness of the underlying UV continuum of a spectrum (where $f_\lambda \propto \lambda^\beta$, and originally measured over $\lambda_{\rm rest} \sim 1300$–2600 Å; D. Calzetti et al. 1994; G. R. Meurer et al. 1999). This quantity has correlations with physical properties within a galaxy, such as star formation history, stellar mass, dust attenuation, Lyman-continuum escape fraction, and metallicity (e.g., S. L. Finkelstein et al. 2012; S. M. Wilkins et al. 2013; R. J. Bouwens et al. 2014; A. B. Rogers et al. 2014; S. Tacchella et al. 2022; M. W. Topping et al. 2022; D. Austin et al. 2023; R. Endsley et al. 2023; A. M. Morales et al. 2024, 2025).

As a result, obtaining accurate measurements of $\beta$ is essential to interpreting the physical conditions within galaxies. Yet, there are multiple approaches to measuring $\beta$, and each comes with distinct advantages and limitations. Ideally, $\beta$ is measured directly from a galaxy's spectrum using a power-law fit across the UV continuum, which was the original use case (D. Calzetti et al. 1994). This provides the most accurate measurement of $\beta$ when studying high-quality spectra. However, observational challenges such as sensitivity limits, spectral resolution, or coverage gaps often hinder its use. Such circumstances require one to seek alternative methods to estimate the UV slope.

[21] NSF Graduate Research Fellow.
[22] NASA Postdoctoral Fellow.

In the absence of high-quality spectra, studies commonly rely on broadband photometry to estimate $\beta$, as it can probe larger numbers of fainter sources. Several photometric methods have emerged, including single-color relations (R. J. Bouwens et al. 2010; S. L. Finkelstein et al. 2010; J. S. Dunlop et al. 2012, 2013; A. B. Rogers et al. 2013), power-law fitting across multiple photometric points (R. J. Bouwens et al. 2012; A. B. Rogers et al. 2013, 2014; F. Cullen et al. 2017; M. W. Topping et al. 2022; F. Cullen et al. 2023; A. M. Morales et al. 2024, 2025; M. W. Topping et al. 2024), and full spectral energy distribution (SED) fitting approaches (S. L. Finkelstein et al. 2012; A. B. Rogers et al. 2013; F. Cullen et al. 2017, 2024; R. Bhatawdekar & C. J. Conselice 2021; S. Tacchella et al. 2022; T. Nanayakkara et al. 2023; A. M. Morales et al. 2024, 2025; C. Papovich et al. 2025). For completeness, we note that F. Cullen et al. (2024) adopt a variant of the power-law technique that also incorporates the Lyman break and Ly$\alpha$ damping wing, enabling simultaneous fitting of redshift and $\beta$. While this differs slightly from the continuum-only power-law fits discussed here, it serves a similar goal of constraining UV slopes from photometric data. While these photometric methods offer flexibility, their accuracy is often affected by redshift uncertainties, filter availability, and signal-to-noise limitations, particularly at higher redshifts. Inaccurately measuring $\beta$ can lead to incorrect physical interpretations of galaxies, such as overestimating or underestimating their dust content or inferring misleading stellar properties (S. M. Wilkins et al. 2011).

The launch of the James Webb Space Telescope (JWST; J. P. Gardner et al. 2006, 2023) opens new opportunities to refine our understanding of rest-frame UV spectral slopes. With its higher sensitivity and broader wavelength coverage, JWST's Near Infrared Camera (NIRCam; M. J. Rieke et al. 2023) and Near Infrared Spectrograph (NIRSpec; P. Jakobsen et al. 2022; T. Böker et al. 2023) instruments allow us to directly observe rest-frame UV emission from galaxies out to $z \sim 9$ and beyond. This presents a timely opportunity to reevaluate how well photometric techniques with NIRCam perform against spectroscopically measured benchmarks observed with NIRSpec.

This study aims to assess three standard photometric methods for deriving $\beta$ and compare them to the UV slope measured directly from JWST low-resolution PRISM spectroscopy. We focus on PRISM mode because its high sensitivity to continuum emission, owing to its broad wavelength coverage, makes it ideal for accurate UV slope measurements. We use a sample of galaxies with precise spectroscopic redshifts and reliable spectra to establish a "ground truth" for $\beta$. We compare each photometrically derived UV slope value to the spectroscopic value and assess which method provides the most reliable result. With this study, we aim to provide a practical guide for observers who rely on photometry when spectra are not available, offering best practices for estimating UV slopes in the era of JWST.

This paper is structured as follows. In Section 2, we describe the data we use from the CEERS and RUBIES surveys and our data reduction process. In Section 3, we discuss our galaxy sample and our methods for photometric power-law, SED, and single-color fitting to derive $\beta$ from photometry. We also discuss our approach to measuring the UV slope directly from the spectra. In Section 4, we describe our findings from comparing the various photometric measures to the spectroscopic values. We present our conclusions in Section 5. We use the Planck Collaboration et al. (2020) cosmology of $H_0 = 67.4\ \mathrm{km\,s^{-1}\,Mpc^{-1}}$, $\Omega_m = 0.315$, and $\Omega_\Lambda = 0.685$, and all magnitudes are given in the AB system (J. B. Oke & J. E. Gunn 1983).

## 2. Data

### 2.1. Spectroscopic Data

We use a combination of NIRSpec/PRISM data from the CEERS (S. L. Finkelstein et al. 2024) and RUBIES (A. de Graaff et al. 2025) surveys. CEERS observed six NIRSpec/PRISM micro-shutter assembly (MSA) pointings in the Extended Groth Strip (EGS) field, each for an effective integration time of 3107 s. We use the internal CEERS collaboration V0.7 NIRSpec reduction. While the full details of these reductions will appear in P. Arrabal Haro et al. (2025, in preparation), we summarize the reductions below.

The NIRSpec data were reduced using the JWST Calibration Pipeline version 1.8.5 (H. Bushouse et al. 2022) and CRDS context `jwst_1029.pmap`. The majority of the reduction used the standard pipeline parameters, with a few exceptions to improve the rejection of cosmic rays and "snowball" artifacts, and better optimize the width of the 2D to 1D spectral extraction apertures to better match the width of the apertures to the sizes of the observed sources.

RUBIES observed 12 NIRSpec/PRISM pointings in the PRIMER-UDS field and 6 pointings in the EGS field, each for an effective exposure time of 2889 s. We use the reduction of this data set presented in A. J. Taylor et al. (2025) in this work, focusing only on sources in the EGS field. Briefly, A. J. Taylor et al. (2025) adopted the same JWST Calibration Pipeline modifications used in the CEERS V0.7 reductions (see above), but used a newer pipeline version (1.13.4) and CRDS context (`1215.pmap`). Despite these minor differences, the two reductions yield consistent spectroscopic results across overlapping sources and fields.

### 2.2. Photometric Data

Our photometric catalog follows a similar process to that described by S. L. Finkelstein et al. (2024); therefore, we direct the reader to their work for more details (it will be described fully in S. L. Finkelstein et al. 2025, in preparation). We use the publicly released CEERS v1.0 NIRCam mosaics, as well as the F090W mosaic from PID 2234 (PI: Bañados). Source detection was performed using SOURCE EXTRACTOR (E. Bertin & S. Arnouts 1996) in dual-image mode, with an inverse-variance weighted combination of the F277W and F356W bands used as the detection image. Photometry for the entire catalog was measured in all available NIRCam filters (F090W, F115W, F150W, F200W, F277W, F356W, F410M, and F444W), along with any available Hubble Space

Telescope (HST)/Advanced Camera for Surveys (ACS) imaging (F435W, F606W, and F814W), generally from CANDELS (N. A. Grogin et al. 2011; A. M. Koekemoer et al. 2011). However, for the purposes of this work, only the NIRCam photometric bands were used. Elliptical Kron apertures were used to extract photometry, with total fluxes derived using a two-step correction: (1) the ratio of fluxes in small and large Kron apertures in F277W, and (2) a simulation-based residual aperture correction. Source identification followed a two-pass approach, combining a "cold" mode to detect bright, extended objects and a "hot" mode to recover faint, compact sources missed in the initial pass (see S. A. Rix et al. 2004). To ensure consistent photometry across bands, images with smaller point spread functions (PSFs) than F277W were convolved with empirically derived kernels. For larger-PSF images, corrections were computed based on F277W images convolved to the broader PSF.

## 3. Methodology

In Section 3.1, we describe our sample of high-redshift galaxies. In Section 3.2, we describe the process of deriving the UV spectral slope from spectroscopy. In Section 3.3, we describe the process of deriving the UV spectral slope from photometric power-law fitting to the observed photometry, SED fitting, and the single-color method.

### 3.1. Galaxy Sample

To obtain our initial sample, we run the entirety of the CEERS catalog through EAZY (G. B. Brammer et al. 2010), wherein we obtain the redshift probability distribution, $P(z)$, and best-fit redshift, $z_a$. We used the default set of 12 "tweak FSPS" templates and included 6 additional templates from R. L. Larson et al. (2023), which provide more flexibility in fitting a broader range of SED shapes, including bluer colors (this process is run in a similar manner to A. M. Morales et al. 2025). We assume a flat redshift prior with respect to luminosity and allow the redshift to range from $z = 0$ to 20. No redshift or magnitude priors were imposed during the fitting process, ensuring the results were not biased by prior assumptions. Redshift cuts were applied only after the fitting was completed.

We select a parent galaxy sample photometrically, designing selection criteria based on the photometric redshift probability distribution functions $P(z)$. Sample selection follows the process outlined in S. L. Finkelstein et al. (2025, in preparation), which extends the selection of S. L. Finkelstein et al. (2024) to the lower redshifts probed here. The initial selection criteria require that a galaxy's redshift is robust if the integral of its redshift probability distribution is

1. $\int_{z_a-0.2(1+z_a)}^{z_a+0.2(1+z_a)} P(z) > 0.5$, and
2. $\int_{z_a-2}^{20} P(z) > 0.7$,

where $z_a$ is the redshift where $\chi^2$ is minimized for the all-template linear combination modes with EAZY (G. B. Brammer et al. 2010). We also follow the signal-to-noise cut criteria outlined in S. L. Finkelstein et al. (2024), where we require sources to have a signal-to-noise ratio $(S/N) > 5.3$ (or $>4.3$) in at least two (or three) of the F115W, F150W, F200W, F277W, F356, or F444W filters, and to show a dropout signature with $S/N < 2.5$ in all bands blueward of Ly$\alpha$, allowing at most one filter with $S/N > 2.0$. We find 4188 galaxies that satisfy these criteria.

We cross-match this photometric sample with CEERS and RUBIES spectroscopic redshift catalogs, focusing on sources with $z_{\rm PEAK} \geqslant 5.5$, where $z_{\rm PEAK}$ is the redshift at the peak of the redshift probability distribution returned from EAZY. We initially select galaxies based on photometric redshifts to reflect typical UV slope measurement conditions in photometry-only studies. To ensure reliable sampling of the UV slope, we retain only sources where photometric and spectroscopic redshifts are in reasonable agreement, filtering out extreme outliers and minimizing redshift-driven biases in the UV slope. Galaxies from both spectroscopic datasets are matched to the photometric catalog using a positional threshold of 0$''$5.

The CEERS EGS spectroscopic catalog (P. Arrabal Haro et al. 2025, in preparation) includes 839 high-confidence redshifts derived through template fitting. After a visual vetting process, the precise redshift values of all the sources presenting emission lines were refined using the `Lime` line fitting code (V. Fernández et al. 2024).[23]

For the RUBIES galaxies in the EGS field, we compile spectroscopic redshifts from two sources, yielding a total of 221 galaxies with confirmed spectroscopic redshifts. First, we download and compile spectroscopic redshifts from the DAWN JWST Archive (DJA).[24] We additionally use an internal emission line detection code to supplement these DJA redshifts. In brief, this code searches for significantly detected emission lines in NIRSpec PRISM spectra and fits a spectroscopic redshift by comparing the velocity separations of the detected lines to a list of strong emission lines typically observed in galaxies. This code is still a prototype. Thus, we visually inspect and verify the fidelity of all spectroscopic redshifts estimated using this routine.

To ensure the spectra are of high enough quality to allow a $\beta$ measurement, we calculated the S/N per-pixel within rest-frame $\lambda = 1500$–$3000$ Å, where we intend to measure the UV slope. While UV slopes are traditionally measured over ∼1300–2600 Å (D. Calzetti et al. 1994), we adopt a slightly redder wavelength range to better suit high-redshift galaxies observed with JWST. This range leverages additional photometric coverage from NIRCam and avoids possible contamination from the Ly$\alpha$ damping wing, which affects the bluer end of the measured spectrum at high redshift (this choice is discussed further in Section 3.2). We omit sources from our analysis whose spectra have >60% of the pixels in this range with $S/N < 2$. This flag also removes any sources with gaps in their spectrum within the region where the UV slope is measured. This cut yields a final sample of 53 galaxies, each photometrically selected and spectroscopically confirmed, to be used in our subsequent analysis (See Figure 1 for the redshift and UV slope distribution of this sample).

#### 3.1.1. Slit-loss Correction

We use the available photometry to derive slit losses by deriving a wavelength-dependent correction curve for each galaxy by comparing observed photometry to "synthetic photometry" computed from the spectra. For each object, we compiled photometric fluxes and uncertainties from the seven JWST/NIRCam broadband filters (F090W, F115W, F150W, F200W, F277W, F356W, and F444W). Synthetic photometric fluxes (or the bandpass-averaged fluxes; BPFs) were then calculated by integrating the spectra over the transmission curves of the same filters. Each transmission curve, $\epsilon(\lambda)$, was interpolated onto the spectral wavelength grid and applied to the appropriate spectral range. In Equation (1), $f_\lambda(\lambda)$ is the spectral flux density and $\epsilon(\lambda)$ is the filter transmission:

$$\mathrm{BPF} = \frac{\int \lambda f_\lambda(\lambda)\epsilon(\lambda)d\lambda}{\int \lambda\epsilon(\lambda)d\lambda} \times [\mathrm{erg\ s^{-1}\ cm^{-2}\ \AA^{-1}}]. \quad (1)$$

[23] https://github.com/Vital-Fernandez/lime

[24] https://s3.amazonaws.com/msaexp-nirspec/extractions/nirspec_graded_v3.html

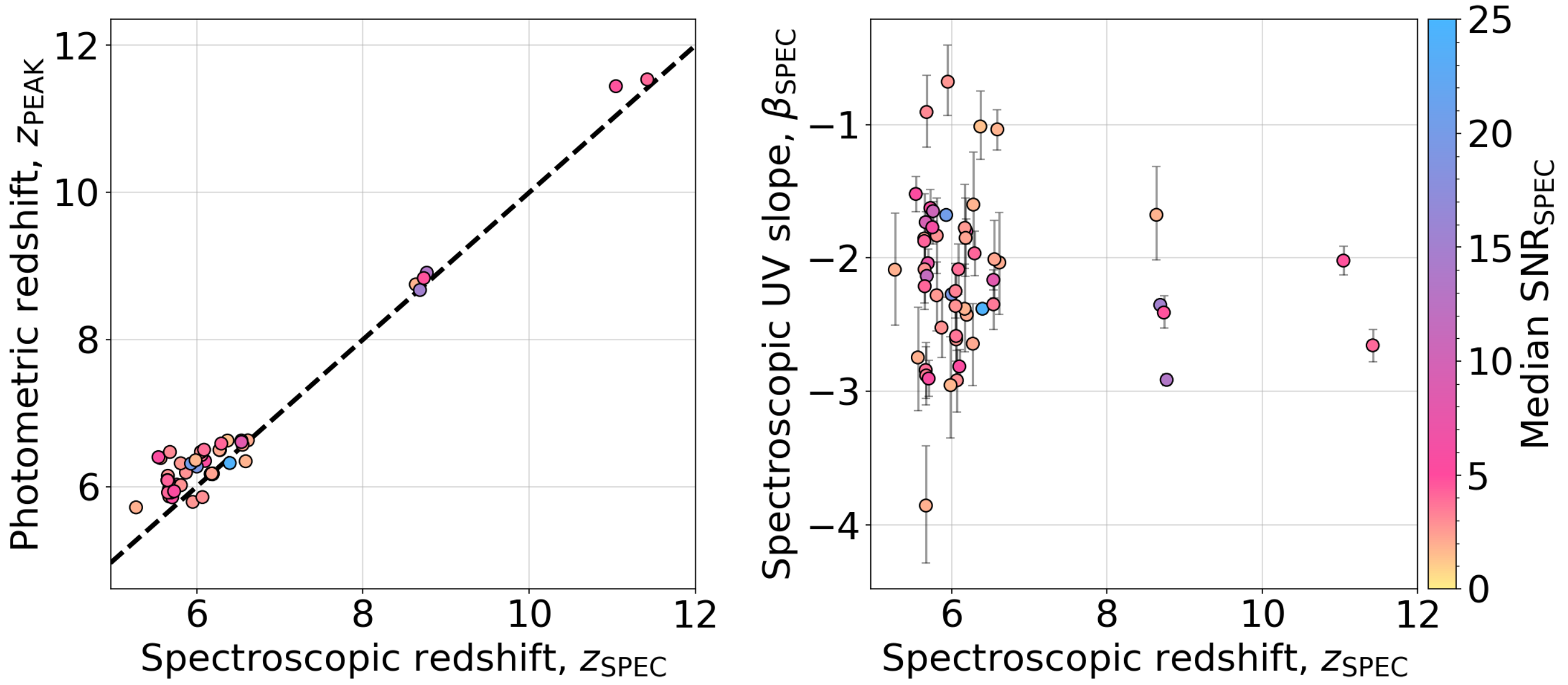

**Figure 1.** The full sample of redshifts and spectroscopic UV slopes for CEERS and RUBIES spectra in this work. (Left) A comparison of spectroscopic redshifts and photometric redshifts from EAZY. The black dashed line is the one-to-one line. Most sources here show good agreement between photometric and spectroscopic measurements for redshift. (Right) Spectroscopic redshift versus the spectroscopic UV slope for the full sample from $z_{\mathrm{SPEC}} \sim 5$–12. Most sources have $\beta_{\mathrm{SPEC}} \sim -2$, typical for UV-selected star-forming galaxies. There is no clear trend in $\beta_{\mathrm{SPEC}}$ with redshift, though the scatter increases particularly where signal-to-noise is lower. A few sources in this sample show very blue slopes, though these measurements carry larger uncertainties. Both subplots are color-coded by median signal-to-noise in the regime where the UV slope is defined (rest-frame $\lambda = 1500$–3000 Å).

To ensure reliability, only filters whose corresponding synthetic photometric fluxes with an $S/N \geqslant 2$ within $\lambda_{\mathrm{rest}} = 1500$–3000 Å were used in the subsequent fitting procedure. For those valid filters, we computed the ratio of observed to synthetic fluxes, yielding a sparse correction function across the photometric wavelength range. We interpolated across gaps in this array and then fit a Chebyshev polynomial of up to third degree to the correction factors, with the degree dynamically set based on the number of valid points ($n - 1$, 3). This polynomial was evaluated across the entire spectral wavelength range, producing a smooth correction curve. We note that this correction procedure is functionally similar to the calibration routine implemented in the SED fitting code BAGPIPES (A. C. Carnall et al. 2018), which also derives a smooth correction curve between spectra and photometry. However, our method isolates this step and achieves comparable results with significantly reduced computational overhead. We defer a complete discussion of BAGPIPES and its use in deriving $\beta$ photometrically to Section 3.3.

The original spectra and their associated uncertainties were then scaled by this curve to generate slit-loss-corrected spectra consistent with broadband photometry. Finally, we recomputed synthetic photometry from the corrected spectra to verify the accuracy of the correction. We compared the original and corrected spectra alongside photometric points and the fitted correction curve for visual inspection. For the remainder of this paper, any analysis discussed uses the observed photometry and the corrected spectra.

### 3.2. Deriving the UV Spectral Slope from JWST PRISM Spectra

For all objects in our final sample of 53 galaxy spectra, we directly measure the UV spectral slope, $\beta_{\mathrm{SPEC}}$, by fitting a power law to each spectrum using the functional form $f_\lambda = A\lambda^\beta$ in linear space across selected rest-frame UV wavelength windows. This approach allows us to properly propagate flux uncertainties, including nondetections and low signal-to-noise measurements. We include all of the D. Calzetti et al. (1994) continuum windows that fall redward of Ly$\alpha$ (i.e., $\lambda > 1500$ Å), effectively encompassing the redder portion of their defined range. Following A. M. Morales et al. (2024), we exclude the bluest Calzetti windows to avoid contamination from Ly$\alpha$ damping wings at high redshift (e.g., P. Arrabal Haro et al. 2023; K. E. Heintz et al. 2024; H. Umeda et al. 2024). We also exclude a new narrow region from $\lambda = 2780$ to 2820 Å to mask potential Mg II emission from stellar populations. Our final fitting range spans the rest-frame interval $\lambda = 1500$–3000 Å, incorporating all valid continuum regions within this range and ensuring robust coverage of the UV slope. This wavelength range is consistent with our photometric fitting methods discussed in Section 3.3, enabling a direct comparison between spectroscopic and photometric measurements of $\beta$.

To better characterize uncertainties in the UV slope and account for covariances with the associated spectra, we employ a Markov Chain Monte Carlo (MCMC) approach using the EMCEE sampler (D. Foreman-Mackey et al. 2013).

**Table 1**
BAGPIPES Priors Defined for This Work, Adapted from A. M. Morales et al. (2024)

| Parameter | Range | Description |
|---|---|---|
| Star-formation History (SFH) Component | | |
| sfh["age"] | (0.002, 13.0) | Age of the galaxy in Gyr |
| sfh["tau"] | (0.1, 14.0) | Delayed decay time in Gyr |
| sfh["metallicity"] | (0.0, 2.5) | Metallicity in units of $Z_{\odot}$ |
| sfh["massformed"] | (4.0, 13.0) | $\log_{10}$ of total mass formed in units of $M_{\odot}$ |
| Dust Component | | |
| dust["type"] | Salim | Shape of the attenuation curve (S. Salim et al. 2018) |
| dust["Av"] | (0.0, 4.0) | Dust attenuation in units of magnitude |
| dust["eta"] | 1.0 | Multiplicative factor on $A_v$ for stars in birth clouds |
| dust["delta"] | (−1.0, 1.0) | Parameter in the Salim dust law that defines the deviation from Calzetti slope |
| dust["B"] | (0.0, 3.0) | 2175 Å (UV extinction) bump strength |
| Nebular Component | | |
| nebular["logU"] | (−4.0, −1.0) | $\log_{10}$ of the ionization parameter |
| nebular["fesc"] | (0.0, 1.0) | Lyman continuum escape fraction |
| Additional Fit Instructions | | |
| fit_instructions["redshift"] | ($z_{\rm PEAK}$ − 1, $z_{\rm PEAK}$ + 1) | Photometric redshift range tested (where $z_{\rm PEAK}$ is the photometric redshift measured by EAZY) |

We fit the power-law model in the form $f_\lambda = 10^A\,\lambda^\beta$ where $A$ is the base-10 logarithm of the flux normalization. Because typical $f_\lambda$ values are very small (e.g., $\sim10^{-20}$ erg s$^{-1}$cm$^{-2}$ Å$^{-1}$), we sample over $A$ rather than the linear normalization to improve numerical stability and aid convergence of the MCMC sampler. We adopt wide, flat priors on both parameters: $-50 < A < 0$ and $-5 < \beta < 5$. The final normalization is computed as $10^A$ during model evaluation, and transformed postrun as needed. The MCMC chains are run with 32 walkers over 5000 steps. These walkers are independent chains that collectively sample the posterior, each taking a path through parameter space to map the distribution. To ensure convergence, we discard the first 3000 steps as burn-in and thin the chains by a factor of 15. We then perform a visual inspection of the chains to verify that the posterior distributions are well sampled. The posterior distributions are marginalized to obtain the medians of the parameters and their 68% confidence intervals. Finally, we evaluate the resulting power-law model and its confidence interval across the observed wavelength range and visually inspect the fit by comparing it to the data.

### 3.3. Deriving the UV Spectral Slope from Different Photometric Approaches

For this work, we compare our spectroscopically derived UV slopes, $\beta_{\rm SPEC}$, to those of three different photometric approaches, where each approach utilizes the photometric redshift, "$z_{\rm PEAK}$" obtained from EAZY (G. B. Brammer et al. 2010): (1) photometric power-law fitting, (2) SED fitting, and (3) single-color fitting. The photometric power-law fitting method mirrors the approach used for fitting $\beta$ from spectra, but instead of fitting the spectrum directly, it fits a power law to the much smaller number of photometric data points. Filters are included if any portion of their transmission curve overlaps with the rest-frame wavelength range $\lambda$ = 1500–3000 Å, analogous to the spectral fitting window. For each selected filter, the flux is assigned to the filter's central wavelength for the fit. Most sources (47 out of 53) are fit using three qualifying filters, while the remaining six only use two. Note that the filter selection in this method differs from that used in the single-color approach, as discussed below. We follow the same procedure for measuring the photometric power-law UV slope, denoted as $\beta_{\rm PL}$, by fitting a power-law model with parameters $\beta_{\rm PL}$ and the scaling factor, A. Both parameters are sampled using the same MCMC framework described in Section 3.2, from which we derive the median values and 68% confidence intervals.

We next measure the UV slope from SED fitting to the observed photometry, $\beta_{\rm SED}$. We run BAGPIPES (A. C. Carnall et al. 2018), a Bayesian SED-fitting code, where photometric data are provided as input along with a suite of priors, and the posterior distributions of galaxy properties and corresponding model spectra are returned. We assume a delayed-tau star formation history, and our priors span an ample enough parameter space to ensure we are not omitting valuable information (see Table 1). We measure $\beta$ directly from these model spectra, following S. L. Finkelstein et al. (2012) and A. M. Morales et al. (2024).

Following A. M. Morales et al. (2024), we incorporate a free parameter added to BAGPIPES, the Lyman continuum escape fraction, $f_{\rm esc}$, to allow for the flexibility of measuring bluer colors within our fitting ($f_{\rm esc}$ was otherwise assumed to be zero). Nebular emission (lines and continuum) is scaled by $(1 - f_{\rm esc})$ to account for the fraction of ionizing photons absorbed and reprocessed by the interstellar medium (ISM), with the remainder escaping without contributing to nebular features. By incorporating this $f_{\rm esc}$ component with other tunable priors, we can allow BAGPIPES to generate a model reaching a minimum blue "floor" of $\beta_{\rm SED} = -3.25$ when $f_{\rm esc} = 1$ (when compared to $\beta_{\rm SED} = -2.44$ when $f_{\rm esc} = 0$). See Figure 1 in A. M. Morales et al. (2024) for the full probability distribution for the UV slope with BAGPIPES.

We measure the SED-fitted UV slope, $\beta_{\rm SED}$, from each of the 1000 posterior model spectra via a power-law fit to all model flux density data points from $\lambda_{rest} = 1500$–$3000$ Å, while excluding any data within the same windows specified above when fitting the observed spectra. We adopt the median and the difference from the 68th percentile from these measurements as our final $\beta_{SED}$ value and uncertainties.

Finally, with single-color fitting, we go about measuring it within two different wavelength ranges. The first utilizes fixed photometric filters closest to rest-frame 1500 and 3000 Å, comparable to the other photometric methods, and we denote it as the single-color (fixed) UV slope, $\beta_{SCF}$. The second finds the two photometric filters whose rest-frame wavelengths are just redward of the Ly$\alpha$ break (rest-frame 1216 Å) for a galaxy at a given redshift, and is denoted as the single-color (break) UV slope, $\beta_{SCB}$. For the two photometric filters used in both instances, flux densities are converted to AB magnitudes, along with their corresponding propagated errors. The UV slope is then derived from the color index (the difference in magnitudes) between the two filters through the following equation:

$$\beta_{SC} = \frac{0.4 \cdot (m_2 - m_1)}{\log_{10}\left(\frac{\lambda_2}{\lambda_1}\right)} - 2, \quad (2)$$

where $m_1$ and $\lambda_1$ correspond to the AB magnitudes and wavelength values for the first filter utilized, and $m_2$ and $\lambda_2$ correspond to the second filter. The uncertainty on the UV slope through the single-color method is computed by error propagation of the previous equation:

$$\sigma_{\beta_{SC}} = \frac{0.4}{\log_{10}\left(\frac{\lambda_2}{\lambda_1}\right)} \cdot \sqrt{\sigma_{m_1}^2 + \sigma_{m_2}^2}. \quad (3)$$

## 4. Results and Discussion

### 4.1. Comparing Spectroscopic and Photometric UV Slopes

In Figure 2, we show two examples of galaxies with corrected NIRSpec PRISM spectra, highlighting the wavelength region ($\lambda = 1500$–$3000$ Å rest-frame) used to fit the UV slope. The top panel (A) shows a galaxy with high median spectroscopic S/N ($\sim$20.5 within $\lambda_{rest} = 1500$–$3000$ Å), where the UV slope estimates, specifically those from spectroscopic fitting ($\beta_{SPEC}$), photometric power-law fitting ($\beta_{PL}$), and SED fitting ($\beta_{SED}$), are consistent within uncertainties and have small error bars. The photometric fit in this case uses two high-S/N photometric points, yielding a similar value compared to the spectral result. The bottom panel (B) shows a galaxy with a lower median spectroscopic S/N ($\sim$4.2 within $\lambda_{rest} = 1500$–$3000$ Å). Here, while $\beta_{SPEC}$ and $\beta_{SED}$ remain in better agreement compared to other techniques, $\beta_{PL}$ shows a bluer slope and larger uncertainty, likely due to reliance on three noisier photometric points with lower S/N. In both examples, the fixed single-color method ($\beta_{SCF}$; utilizing the two photometric filters closest to rest-frame wavelengths 1500 and 3000 Å) yields a UV slope that is roughly comparable but still offset from the more robust methods. The break-based single-color method ($\beta_{SCB}$), which uses the first two filters redward of the Lyman break, shows the largest deviation from $\beta_{SPEC}$ in both cases, reflecting its sensitivity to filter spacing and lower reliability.

Figure 3 compares each photometric UV slope method to the spectroscopic benchmark across the full sample. Residuals, defined as $\Delta\beta = \beta_{PHOT} - \beta_{SPEC}$, assess how well each method recovers the true UV slope. To account for measurement uncertainties when comparing UV slope estimates, we calculate normalized residuals, defined as the difference between the photometric and spectroscopic slopes divided by the combined uncertainty from both methods. Specifically, the combined uncertainty is estimated as the quadrature sum of the errors from each method, defined as: $\sigma_{RES} = \sqrt{\sigma_{PHOT}^2 + \sigma_{SPEC}^2}$, where $\sigma_{PHOT}$ and $\sigma_{SPEC}$ are the respective uncertainties on the photometric and spectroscopic $\beta$ measurements. The normalized residual is then: $\Delta\beta_{NORM} = \Delta\beta/\sigma_{RES}$.

In both instances, both $\beta_{PL}$ and $\beta_{SED}$ show overall good agreement with $\beta_{SPEC}$ with small residual scatter (e.g., mean $\Delta\beta = 0.027$ for $\beta_{PL}$, $-0.003$ for $\beta_{SED}$) and low normalized residuals (mean $\Delta\beta_{NORM} = 0.315$ and 0.126, respectively). The $\beta_{SCF}$ method performs modestly worse, showing slightly larger residuals (mean $\Delta\beta = 0.284$ and mean $\Delta\beta_{NORM} = 1.246$). The $\beta_{SCB}$ method shows the largest scatter and systematic bias (mean $\Delta\beta = -0.851$ and mean $\Delta\beta_{NORM} = -1.679$), consistent with expectations given its limited wavelength baseline. We include $\beta_{SCB}$ for historical context, as it was commonly employed in studies using HST data prior to the advent of JWST (R. J. Bouwens et al. 2009, 2014; S. L. Finkelstein et al. 2012).

### 4.2. Recommendations for Best Practices

Photometric power-law fitting ($\beta_{PL}$) can yield UV slopes close to those derived from spectroscopy ($\beta_{SPEC}$), especially when at least 2–3 photometric points in the UV regime have high S/N. This makes $\beta_{PL}$ a useful and computationally cheap technique when spectra are unavailable, especially with JWST/NIRCam, which offers deeper and broader rest-UV coverage than previous facilities like HST. Importantly, $\beta_{PL}$ does not show strong systematic biases in our sample and generally returns robust values when photometry is of sufficient quality.

SED fitting ($\beta_{SED}$), while more model-dependent, offers a powerful alternative when sources have limited or noisier photometry. As shown in Figure 2(B), $\beta_{SED}$ more closely agrees with $\beta_{SPEC}$ than $\beta_{PL}$ when photometric data are sparse or noisy. This is because SED fitting incorporates the full photometric SED and models the galaxy, effectively inferring a physical spectral shape, mitigating the impact of individual low-S/N bands. Thus, for faint galaxies or those with limited UV coverage, particularly at $z \gtrsim 7$, SED fitting may be a good comparison or sanity check to photometric power-law fitting and perhaps provide further physical information on a galaxy, assuming the models are a good fit to the data. However, if the models do not capture the true range of galaxy properties, biases may arise. We therefore use residuals as a check on the reliability of all methods.

That said, both $\beta_{PL}$ and $\beta_{SED}$ exhibit mild biases in some cases. For $\beta_{PL}$, residuals tend to deviate from the zero line in galaxies with fewer (>3) and/or noisier UV bands. For $\beta_{SED}$, the discrepancy may result from limitations in the stellar model grids one utilizes (this work uses BC03 stellar model grids, by default; G. Bruzual & S. Charlot 2003). This can underpredict very blue continua due to assumptions in the input stellar tracks or dust attenuation curves. The redshift assumed for photometric model fitting can also differ slightly from the spectroscopic redshift. Consequently, the UV slope derived from photometric SED fitting may probe a slightly different portion of the spectrum: a photometric model at a slightly higher redshift will map observed bands to slightly bluer rest-frame wavelengths than the spectroscopic measurement, potentially introducing small offsets in the derived $\beta$.

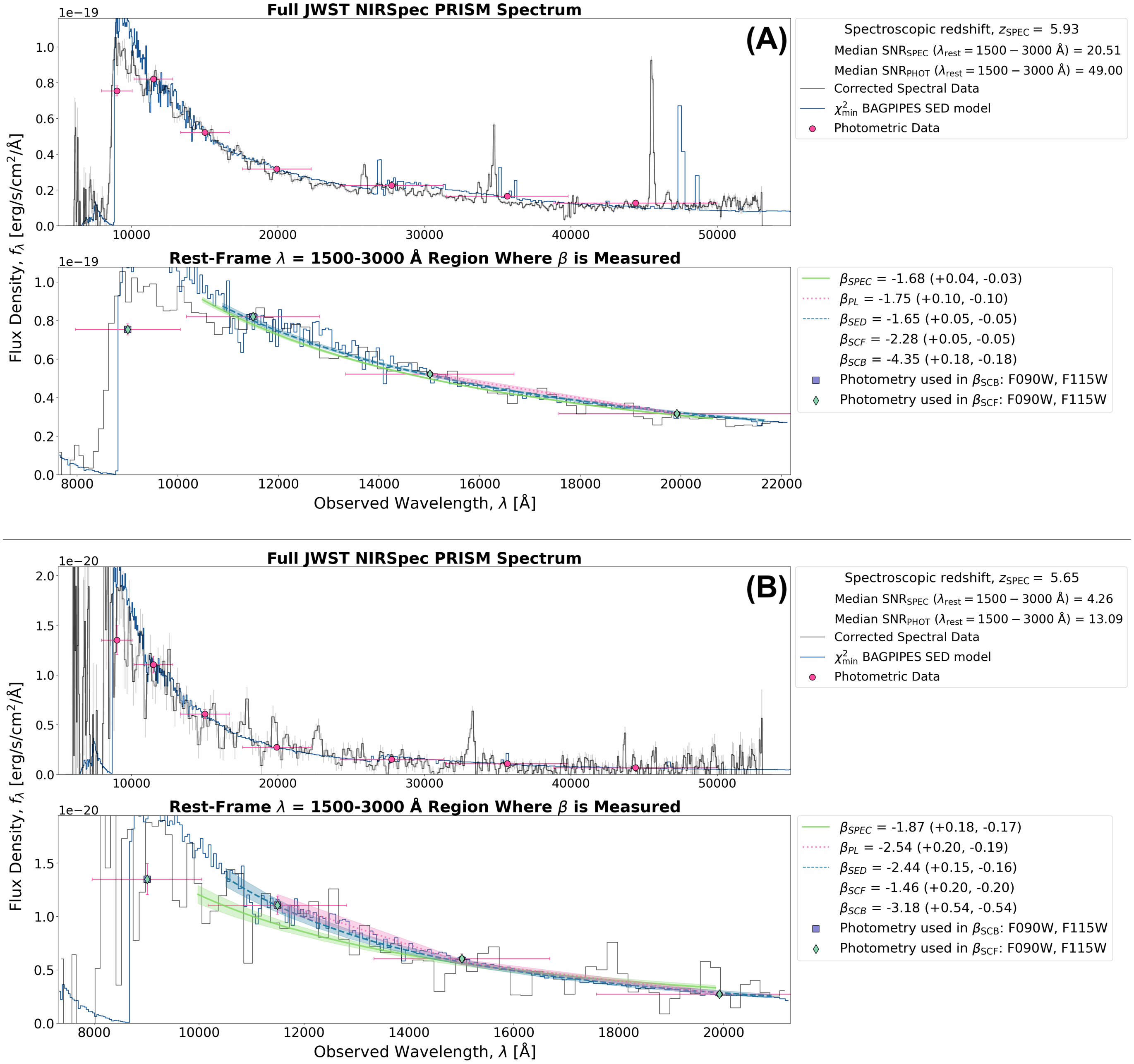

**Figure 2.** Two examples of corrected PRISM spectra, with UV $\beta$ slopes derived from spectra power-law fitting (green), photometric power-law fitting (pink), SED fitting with BAGPIPES (blue), and single-color fitting (the filters used for $\beta_{\rm SCB}$ are shown as purple square markers, and $\beta_{\rm SCF}$ are shown as green diamond markers). The median uncertainty for each $\beta$ is derived from the upper and lower error bars ($1\sigma$ difference from the median of each $\beta$ distribution). The blue SED is the least $\chi^2$ model derived from BAGPIPES when fit to a source's photometry and photometric redshift to see how it compares to the observed PRISM spectrum (any offset in the emission lines and Lyman break is a result of the minor difference between spectroscopic and photometric redshift measurements). The top panel (A) shows a high-S/N (∼20 measured from the spectrum across $\lambda_{\rm rest}$ = 1500–3000 Å) galaxy spectrum where all methods (except the single-color (break) method, $\beta_{\rm SCB}$) yield relatively consistent $\beta$ values with small uncertainties. With lower panel (B), where the median spectroscopic S/N is low (∼4), uncertainties increase for all methods, and the photometric power-law fit ($\beta_{\rm PL}$), which relies on three noisier data points in this case, begins to diverge. In contrast, the SED-fitted UV slope ($\beta_{\rm SED}$) remains in slightly better agreement with the spectroscopic UV slope result, suggesting it may provide a more comparable estimate than $\beta_{\rm PL}$ in low-S/N regimes. The single-color method again deviates, showing its limitations in such conditions.

We also tested whether subtracting strong nebular emission lines (e.g., H$\beta$ and [O III]) from the photometry and fitting stellar-only components affected the recovered UV slopes from SED fitting. The results were not significantly different from the original fits, and in some cases the stellar-only fits yielded marginally worse agreement with the spectroscopic slopes. This suggests that nebular emission modeling has, perhaps, little impact on the inferred UV slope, with differences instead driven by photometric quality and redshift uncertainties.

Another potential source of systematic bias in UV slope measurements is the presence of the 2175 Å dust attenuation feature, often referred to as the "2175 Å bump" (M. Kriek &

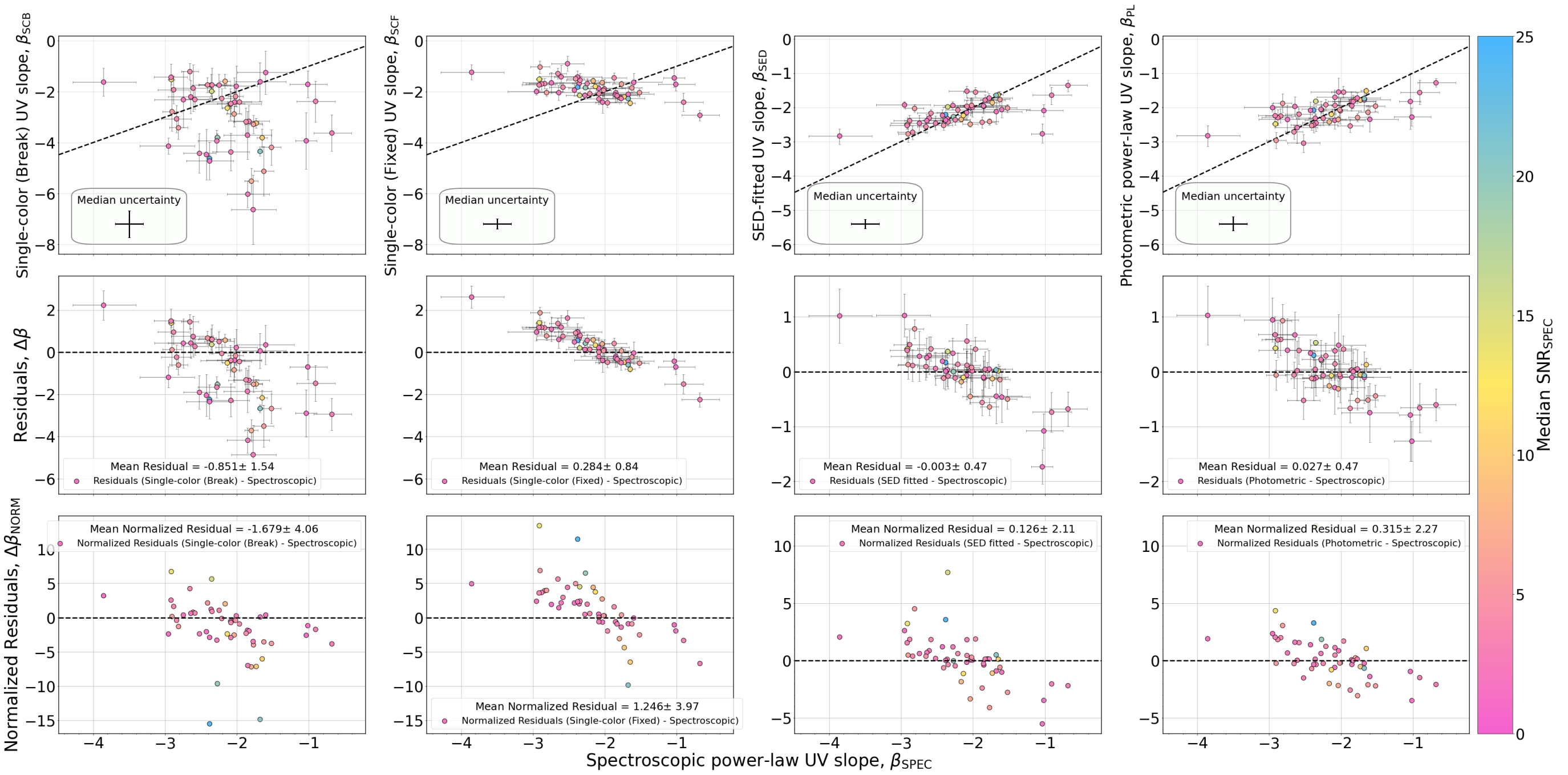

**Figure 3.** Comparison of various photometric UV slope estimation techniques with spectroscopic UV slope measurements. The methods shown include: (1) single-color fitting using the first two filters redward of the Lyman break, (2) single-color fitting at fixed rest-frame wavelengths closest to $\lambda_{\rm rest} = 1500$ and 3000 Å, (3) SED fitting, and (4) photometric power-law fitting. Each technique is compared to spectroscopic power-law-fitted UV slopes. The residuals for each method are displayed in the second row of subplots and are color-coded by the median signal-to-noise ratio (S/N) in the spectral range where $\beta$ is measured spectroscopically ($\lambda_{\rm rest} = 1500$–3000 Å). In the top panels, the black dashed lines indicate the one-to-one relationship between spectroscopic and photometric $\beta$ values. Insets display the median uncertainties for each photometric technique and spectroscopic fitting. The second row shows the mean residuals, defined as $\Delta\beta = \beta_{\rm PHOT} - \beta_{\rm SPEC}$, using only the median values per bin. Among all methods, photometric power-law fitting and SED fitting achieve the most accurate recovery of the spectroscopic UV slope, which is assumed to be the "true" $\beta$. Single-color methods perform the worst, with the highest residuals. However, the fixed-wavelength single-color method better approximates realistic UV slopes than the break-based single-color method, yielding results more comparable to SED and photometric power-law fits. The bottom row presents normalized residuals, $\Delta\beta_{\rm NORM} = \Delta\beta/\sigma_{\rm RES}$, which account for the combined photometric and spectroscopic uncertainties. These plots confirm the same trends: photometric power-law fitting and SED fitting offer the most consistent recovery of $\beta$, while single-color (break) fitting performs the worst.

C. Conroy 2013; S. Salim et al. 2018). If present in a galaxy's extinction curve, this feature can distort the observed UV continuum shape, especially in systems with moderate dust attenuation (i.e., nonnegligible $A_{\rm v}$). The impact of this feature depends on both the amount of dust and, potentially, the orientation of the galaxy relative to the line of sight, as inclined systems may exhibit more substantial effective attenuation. To account for this feature, with BAGPIPES, we incorporate the Salim dust model (S. Salim et al. 2018), a flexible dust model based on a large sample of local galaxies spanning a wide range of star formation activity. This dust law allows for a broad range of slopes and includes a variable 2175 Å UV bump. The slope of the curve correlates primarily with galaxy optical opacity, becoming shallower in more dust-rich systems. Other photometric and spectroscopic methods do not explicitly account for this feature in our measurements of the UV slope.

In contrast, the single-color methods ($\beta_{\rm SCB}$ and $\beta_{\rm SCF}$) are the least reliable across our sample. $\beta_{\rm SCB}$, which takes filters redward of the Lyman break, shows the largest scatter and most significant deviation from spectroscopic results. Whereas $\beta_{\rm SCF}$, which utilizes photometric filters closest to rest wavelengths 1500 and 3000 Å, yields results more consistent with that of photometric power-law fitting and SED fitting, though still less accurate overall. While the simplicity and minimal data requirements of single-color methods make them appealing for very high-redshift or shallow observations, this comes at the cost of both precision and reliability, particularly for galaxies with complex SEDs or limited UV photometric coverage.

Finally, the residuals shown in Figure 3 reveal an anticorrelation across all photometric methods when compared to spectroscopic measurements of the UV slope. While biases or restrictions from each of the photometric methods may contribute to these trends, as mentioned above, we also note the possibility that the quality of PRISM data may also contribute to this. For each set of residuals versus spectroscopic slope, we plot the median residual value along with the propagated uncertainties. Focusing on the SED-fitting and photometric power-law methods, we find that the highest-S/N cases yield average residuals near zero and exhibit low uncertainties in both the residuals and spectroscopic slopes. In contrast, lower-S/N sources not only amplify the anticorrelation trend but also display larger uncertainties in both $\beta_{\rm SPEC}$ and the residuals. We additionally measured the standard deviation of the residual distribution as a function of spectroscopic S/N: for both methods, low-S/N sources have a typical spread of $\Delta\beta \sim 0.5$, whereas the highest-S/N sources show a much tighter distribution with $\Delta\beta < 0.1$.

While our methodology and recommendations are primarily developed and validated for high-redshift galaxies ($z \sim 5$–12) observed with JWST, many of the best practices outlined here also apply to UV slope measurements across a broader range of redshifts, provided that similar data quality and wavelength coverage are available. Based on these results, we recommend the following:

1. Photometric power-law fitting can be applied whenever multiple UV bands are available, including cases with only two filters. Note that while this method can operate with just two bands, it differs from single-color (fixed and break) fitting in how those bands are selected and used. Single-color fitting (fixed) uses the pair of filters near rest-frame 1500 and 3000 Å, whereas photometric power-law fitting fits a slope across any available filters within the UV window ($\lambda_{\rm rest}$ = 1500–3000 Å), depending on the photometric coverage for each galaxy. Thus, the two approaches may draw on different subsets of photometric data points, even when using only two points, and are conceptually distinct.
2. SED fitting provides a comparably robust estimate of the UV slope and may be particularly useful in situations with sparse or noisy photometry (e.g., at high redshift). In such cases, SED fitting can leverage prior information from full SED models, making it a practical alternative to photometric power-law fitting.
3. Avoid relying on single-color fitting from the break except as a rough diagnostic when no other method is feasible.

## 5. Conclusions

Using data from the JWST CEERS and RUBIES surveys, we compare different methods for measuring the rest-frame UV spectral slope of a sample of high-redshift galaxies at $z \sim 5$–12. We measure the UV spectral slope directly from NIRSpec PRISM spectra, as well as through three different approaches using photometry: photometric power-law fitting, SED fitting, and single-color fitting. We compare our photometric UV slopes to the measured spectroscopic UV slopes and determine which photometric approach is ideal when spectra are unavailable. With these comparisons, we reach the following conclusions:

1. For the galaxies in our sample, both photometric power-law fitting and SED fitting provide comparably robust estimates of the UV slope when multiple UV photometric bands are available. Photometric power-law fitting shows little systematic bias and generally recovers spectroscopic slopes well, especially when 2–3 or more high S/N bands are present, making it a simple and computationally efficient option. In contrast, SED fitting is more model-dependent but proves to be valuable when photometry is sparse or noisy, since it leverages the full SED to mitigate the impact of low S/N bands. However, the uncertainties from SED fitting may be artificially reduced if the models do not span the true galaxy properties. Both methods exhibit mild systematic biases at the lowest S/Ns, where residuals tend to showcase that the median photometric UV slope values are further from the spectroscopic UV slope values (though this could also be a result of the quality of the PRISM data itself for this work). The single-color fitting technique remains the least reliable, with the largest scatter and weakest correlation to spectroscopic UV slopes.
2. When measuring $\beta$ via the spectroscopic power-law method for the entire sample, we obtain an average UV slope, $\beta_{\rm SPEC} = -2.13^{+0.45}_{-0.58}$. We compare this to the average photometric UV slopes for the entire sample, where $\beta_{\rm SED} = -2.15^{+0.33}_{-0.37}$ via SED fitting, $\beta_{\rm PL} = -2.08^{+0.32}_{-0.39}$ via photometric power-law fitting, $\beta_{\rm SCF} = -1.85^{+0.35}_{-0.37}$ via single-color fitting at fixed rest-wavelengths (1500 and 3000 Å), and $\beta_{\rm SCB} = -2.65^{+0.91}_{-1.72}$ via single-color fitting redward of the Lyman break. These average blue UV slopes are indicative of young, relatively dust-poor stellar populations, as expected for typical star-forming galaxies within the redshift range of our study, $z \sim 5$–12.

These results highlight the utility of JWST's extensive photometric capabilities for reliably estimating UV spectral slopes in high-redshift galaxies, particularly when spectroscopic observations are unavailable or of insufficient quality. As photometric samples continue to grow in size and redshift coverage, the ability to accurately recover UV slopes using reliable techniques becomes increasingly essential for constraining dust attenuation, star formation histories, and stellar population properties in the early Universe. Our comparison to spectroscopic benchmarks provides a framework for calibrating and selecting the most robust photometric approaches moving forward. We hope this work serves as a practical guide for observers who must rely on photometry, enabling them to make more informed choices about which method to apply based on data quality, filter coverage, and scientific goals.

## Acknowledgments

We acknowledge that the location where this work took place, the University of Texas at Austin, which sits on indigenous land. The Tonkawa lived in central Texas, and the Comanche and Apache moved through this area. We pay our respects to all the American Indian and Indigenous Peoples and communities who have been or have become a part of these lands and territories in Texas, on this piece of Turtle Island.

We thank the RUBIES team (including PIs A. de Graaff and G. Brammer) for delivering this program and sharing the data. A.M.M. acknowledges support from the National Science Foundation Graduate Research Fellowship Program under grant No. DGE 2137420. Any opinions, findings, conclusions, or recommendations expressed in this material are those of the author(s) and do not necessarily reflect the views of the National Science Foundation. A.M.M. and S.L.F. acknowledge support from NASA via STScI JWST-ERS-1345 and JWST-GO-2079.

The JWST data presented in this article were obtained from the Mikulski Archive for Space Telescopes (MAST) at the Space Telescope Science Institute. The specific observations analyzed can be accessed via doi:10.17909/z7p0-8481 and doi:10.17909/0eta-ym85.

*Software*: `IPython` (F. Pérez & B. E. Granger 2007), `matplotlib` (J. D. Hunter 2007), `NumPy` (S. Van Der Walt et al. 2011), `SciPy` (T. E. Oliphant 2007), `Astropy` (T. P. Robitaille et al. 2013; Astropy Collaboration et al. 2018, 2022), `Bagpipes` (A. C. Carnall et al. 2018), `Emcee` (D. Foreman-Mackey et al. 2013), `EAZY` (G. B. Brammer et al. 2010), `scikit-learn` (F. Pedregosa et al. 2011).

## ORCID iDs

Alexa M. Morales https://orcid.org/0000-0003-4965-0402
Steven L. Finkelstein https://orcid.org/0000-0001-8519-1130
Pablo Arrabal Haro https://orcid.org/0000-0002-7959-8783

Micaela B. Bagley https://orcid.org/0000-0002-9921-9218
Antonello Calabrò https://orcid.org/0000-0003-2536-1614
Óscar A. Chávez Ortiz https://orcid.org/0000-0003-2332-5505
Kelcey Davis https://orcid.org/0000-0001-8047-8351
Mark Dickinson https://orcid.org/0000-0001-5414-5131
Eric Gawiser https://orcid.org/0000-0003-1530-8713
Mauro Giavalisco https://orcid.org/0000-0002-7831-8751
Nimish P. Hathi https://orcid.org/0000-0001-6145-5090
Michaela Hirschmann https://orcid.org/0000-0002-3301-3321
Jeyhan S. Kartaltepe https://orcid.org/0000-0001-9187-3605
Anton M. Koekemoer https://orcid.org/0000-0002-6610-2048
Arianna S. Long https://orcid.org/0000-0002-7530-8857
Ray A. Lucas https://orcid.org/0000-0003-1581-7825
Fabio Pacucci https://orcid.org/0000-0001-9879-7780
Casey Papovich https://orcid.org/0000-0001-7503-8482
Nor Pirzkal https://orcid.org/0000-0003-3382-5941
Anthony J. Taylor https://orcid.org/0000-0003-1282-7454
Alexander de la Vega https://orcid.org/0000-0002-6219-5558
Stephen M. Wilkins https://orcid.org/0000-0003-3903-6935
L. Y. Aaron Yung https://orcid.org/0000-0003-3466-035X